\documentclass[letter]{aa}
\usepackage{graphicx}
\usepackage{txfonts}
\usepackage{natbib}

\newcommand{\numax}{\mbox{$\nu_\mathrm{max}$}}

\begin{document}
%

    \title{Probing populations of red giants in the galactic disk with CoRoT\thanks{The CoRoT space mission was developed and is operated by the French space agency CNES, with the participation of ESA's RSSD and Science Programmes, Austria, Belgium, Brazil, Germany, and Spain.} }


   \author{A. Miglio\inst{1}\fnmsep\thanks{Postdoctoral Researcher, Fonds de la Recherche Scientifique - FNRS, Belgium}
          \and
          J. Montalb\'an\inst{1}
          \and
          F.~Baudin\inst{2}
          \and
          P. Eggenberger\inst{1,3}
          \and
          A. Noels\inst{1}
          \and
          S. Hekker\inst{4,5,6}
          \and
          J. De Ridder\inst{5}
          \and
          W. Weiss\inst{7}
          \and
          A. Baglin\inst{8}
          }

   \institute{Institut d'Astrophysique et de G\'eophysique de l'Universit\'e de Li\`ege,
All\'ee du 6 Ao\^ut, 17 B-4000 Li\`ege, Belgium
              \and
               Institut d'Astrophysique Spatiale (IAS), B\^atiment 121, F-91405, Orsay Cedex, France
         \and
             Observatoire de Gen\`eve, Universit\'e de Gen\`eve, 51 chemin des Maillettes, 1290 Sauverny, Switzerland
         \and
             Royal Observatory of Belgium, Ringlaan 3, 1180 Brussels, Belgium
         \and
             Instituut voor Sterrenkunde, K.U. Leuven, Celestijnenlaan 200D, 3001 Leuven, Belgium
         \and
            School of Physics and Astronomy, University of Birmingham, Edgbaston, Birmingham B15 2TT, United Kingdom
         \and
         Institute for Astronomy, University of Vienna, T\"urkenschanzstrasse 17, A-1180 Vienna
         \and
         LESIA, UMR8109, Universit\'e Pierre et Marie Curie, Universit\'e Denis Diderot, Observatoire de Paris, 92195 Meudon,
France  }

   \date{Received; accepted}


  \abstract
   {The detection with CoRoT of solar-like oscillations in nearly 800 red giants in the first 150-days long observational run paves the way for detailed studies of populations of galactic-disk red giants.} 
   {We investigate which information on the observed population can be recovered by the distribution of the observed seismic constraints: the frequency of maximum oscillation power (\numax) and the large frequency separation ($\Delta\nu$).}
   {We propose to use the observed distribution of \numax\ and of $\Delta\nu$ as a tool for investigating the properties of galactic red-giant stars through comparison with simulated distributions based on synthetic stellar populations.}
    {We can clearly identify the bulk of the red giants observed by CoRoT as red-clump stars, i.e. post-flash core-He-burning stars. The distribution of \numax\ and of $\Delta\nu$ gives us access to the distribution of the stellar radius and mass, and thus represent a most promising probe of the age and star formation rate of the disk, and of the mass-loss rate during the red-giant branch.}
  {CoRoT observations are supplying seismic constraints for the most populated class of He-burning stars in the galactic disk. This opens a new access gate to probing the properties of red-giant stars that, coupled with classical observations, promises to extend our knowledge of these advanced phases of stellar evolution and to add relevant constraints to models of composite stellar populations in the Galaxy.}


   \maketitle
\section{Introduction}
\label{sec:corot}
The CoRoT satellite \citep{Baglin06a} is providing high-precision, long, and uninterrupted photometric monitoring of thousands of stars in the fields primarily dedicated to the search for exoplanetary systems (EXOField). These observations reveal a wealth of information on the  properties of solar-like oscillations in a large number of red giants \citep[see ][]{DeRidder09}, i.e. their oscillation frequency range, amplitudes, lifetimes, and nature of the modes (radial as well as non-radial modes were detected). The goldmine of information represented by this detection is currently being exploited.

As described in detail by \citet{Hekker09}, about 800 solar--like pulsating red giants were identified in the field of the first CoRoT 150-day long run in the direction of the galactic centre (LRc01).
\citet{Hekker09} searched for the signature of a power excess due to solar-like oscillations in all stars brighter than $m_{\rm V}$=15 in a frequency domain up to 120 $\mu$Hz. They find that the frequency corresponding to the maximum oscillation power (\numax) and the large frequency separation of stellar pressure modes ($\Delta\nu$) are non-uniformly distributed.
The distributions of \numax\ and $\Delta\nu$ shows single dominant peaks located at $\sim30$ $\mu$Hz and $\sim4$ $\mu$Hz, respectively. The corresponding histograms are also shown in the lower panels of Figs.\ref{fig:numax} and \ref{fig:dnu}. We note, however, that the detection of solar-like pulsations as in the low-frequency end ($\nu_{\rm max} \lesssim 20$ $\mu$Hz) is compromised by the long-period trends, the activity, and granulation signal. We should therefore regard the observed distribution at the lowest frequencies as strongly biased \citep[see ][]{Hekker09}.
%

In this work we focus on the simplest and most robust seismic constraints provided by the observations, \numax\ and $\Delta\nu$, and on the information they supply on the stellar parameters.
As suggested by \citet{Brown91}, \numax\ is expected to scale as the acoustic cutoff frequency of the star, which defines an upper limit to the frequency of acoustic oscillation modes. Following \citet{Kjeldsen95} and rescaling to solar values, \numax\ can be expressed as
\begin{equation}
\nu_{\rm max}=\frac{M/M_\odot}{(R/R_\odot)^2\sqrt{T_{\rm eff}/5777 K}}\ \rm3050\ \mu Hz.
\label{eq:numax}
\end{equation}

The large frequency separation of stellar p modes, on the other hand, is well known to be inversely proportional to the sound travel time in the stellar interior, and is therefore related to the mean density of the star by the following expression \citep[see e.g. ][]{Kjeldsen95}:

\begin{equation}
\Delta\nu=\sqrt{\frac{M/M_\odot}{(R/R_\odot)^3}}\ 134.9\ \rm \mu Hz.
\label{eq:dnu}
\end{equation}

In the following sections we investigate the properties of the underlying stellar population that determine the observed \numax\ and $\Delta\nu$ distributions, and whether the presence of a single peak in \numax\ and $\Delta\nu$ is compatible with current models of red-giant populations in the galactic disk.


\section{Stellar population synthesis}
\label{sec:pop}
To simulate the composite stellar population observed in the LRc01 of CoRoT's EXOField, we used the  stellar population synthesis code {\sc trilegal}\footnote{\texttt{http://stev.oapd.inaf.it/cgi-bin/trilegal}} \citep{Girardi05} designed to simulate photometric surveys.
In {\sc trilegal}, several model parameters (such as the star formation history and the morphology of different galactic components) were calibrated to fit Hipparcos data \citep{Perryman97}, as well as star counts from deeper photometric surveys, such as 2MASS, \citep{Cutri03}, CDFS \citep{Arnouts01}, and DMS \citep{Osmer98}.

In our {\sc trilegal} simulation (T1), we adopted the standard parameters for the four components of the Galaxy (halo, bulge, thin and thick disk). We simulated the stellar population in the sky area observed by CoRoT during LRc01 (3.9 deg$^2$ centred at $\alpha=290.89^\circ$, $\delta=0.46^\circ$) and considered stars in the observed magnitude range ($11 \lesssim m_{\rm V} \lesssim 15$). This population turns out to be dominated by thin-disk stars.
   \begin{figure}
   \centering
   \includegraphics[width=\hsize]{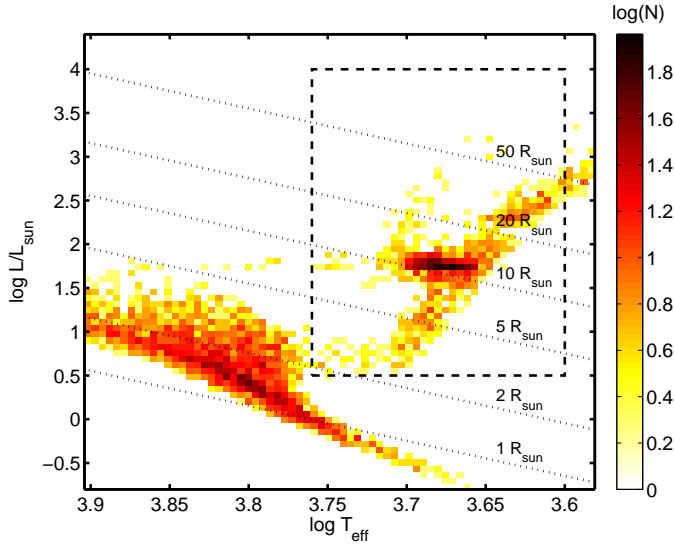}
      \caption{Theoretical Hess diagram of the population T1 simulated with {\sc trilegal}. The dashed rectangle delimits the location in the HR diagram of the population of G-K giants considered in this work. The bar on the right shows the number of stars per $\log{L/L_\odot}-\log{T_{\rm eff}}$ bin on a colour-coded logarithmic scale.}
         \label{fig:hess}
   \end{figure}

The default star formation rate in the thin disk is assumed to be constant over the past 9 Gyr, and the age-metallicity relation is taken from \citet{Rocha00},
and the \citet{Chabrier01} lognormal initial mass function is assumed. 
An exhaustive description of the evolutionary tracks used in the simulations is given by \citet{Girardi05} and references therein. Here for the discussion that follows, we just recall that stellar models are computed assuming \citet{Reimers75} mass-loss rate with an efficiency parameter $\eta=0.4$.

An HR diagram showing the number density of stars (Hess diagram) of the synthetic population is presented in Fig. \ref{fig:hess}, where the main sequence and the red clump (RC) clearly appear as the most populated areas.
   \begin{figure}
   \centering
   \includegraphics[width=.95\hsize]{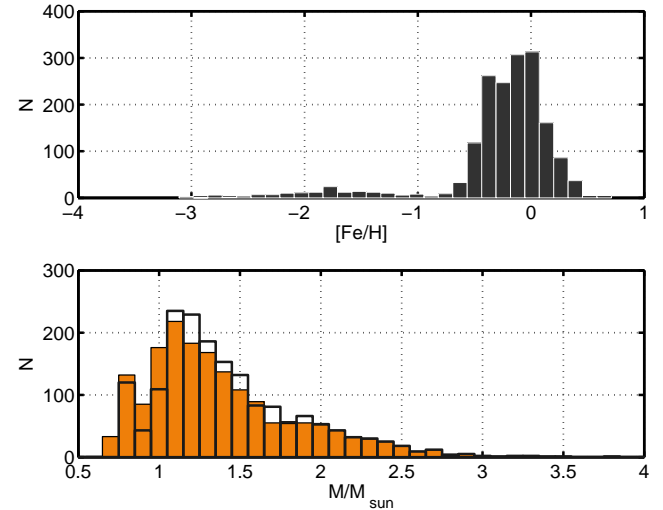}

      \caption{\emph{Upper panel:} Histogram showing the [Fe/H] distribution of the population of G-K giants simulated with {\sc trilegal}. Stars with [Fe/H] $\lesssim$ -1 originate from the (negligible) contribution of the galactic halo. \emph{Lower panel:} The mass distribution of the population is represented by full bars, whereas empty bars illustrate the mass distribution of the progenitors of G-K giants.
              }
         \label{fig:MZ}
   \end{figure}
We henceforth restrict our interest to the population of G-K giants (Fig. \ref{fig:hess}). In the HR diagram (see Fig. \ref{fig:hess}), the distribution of giants is not uniform and predominantly peaked in the red clump, near iso-radii lines corresponding to $R \simeq 10$ $R_\odot$. 
The mass and metallicity distribution of the population is shown in Fig. \ref{fig:MZ}.
We notice that the mass distribution is significantly affected by the mass loss rate assumed during the ascent on the red-giant branch (RGB), which is relevant for stars with initial mass $M_{\rm i} \lesssim 2 M_\odot$. The metallicity distribution of the stars in the population shows a small contribution from halo stars with $\rm [Fe/H] \lesssim -1.0$, and a mean close to the solar metallicity ($\left<\rm[Fe/H]\right>=-0.15$) with a standard deviation of 0.25.

\section{Observed vs. simulated \numax\ and $\Delta\nu$ distributions}
\label{sec:results}
We now simply evaluate, by means of Eqs. \ref{eq:numax} and \ref{eq:dnu}, the theoretical \numax\ and $\Delta\nu$ distributions based on the properties ($R$, $M$, $T_{\rm eff}$) of the synthetic population presented in the previous section.
Among all the stars in the population, we select G-K giants ($0.5 < \log{L/L_\odot} < 4$ and $3.6 < \log{T_{\rm eff}} < 3.76$, see Fig. \ref{fig:hess}). As discussed in Sect. \ref{sec:corot}, we should restrict the comparison to \numax\ $\gtrsim 20$ $\mu$Hz.

\begin{figure}
   \centering
   \includegraphics[width=.9\hsize]{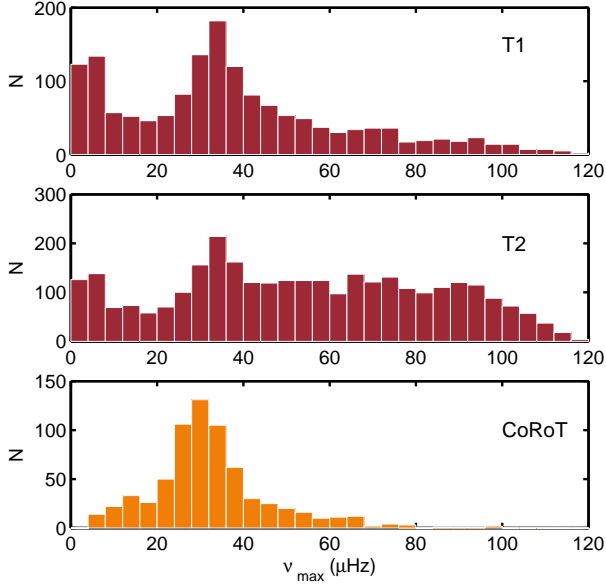}
   \caption{Histogram showing the comparison between the \numax\ distribution of the observed (\emph{lower panel}) and simulated populations of red giants: T1 (\emph{upper panel}) and T2 (\emph{middle panel}), where the effect of a recent burst in the star formation rate is considered (see Sect. \ref{sec:discu}).
              }
         \label{fig:numax}
   \end{figure}

The \numax\ distribution we obtain for T1 is reported in the upper panel of Fig. \ref{fig:numax}. It is clear that the dominant maximum in the observed histogram (lower panel of Fig. \ref{fig:numax}) can be reproduced qualitatively by the stellar population model.

   \begin{figure}
   \centering
   \includegraphics[width=.9\hsize]{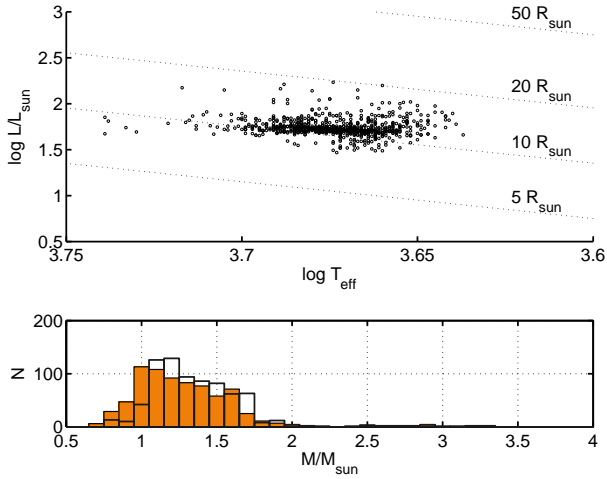}
   \caption{\emph{Upper panel:} HR diagram showing the position of stars in the population T1 with 20 $\mu$Hz $< \nu_{\rm max} < 50$ $\mu$Hz. The radius distribution of these stars has a mean of 11.2 $R_\odot$ with a standard deviation of 1.5 $R_\odot$.
           \emph{Lower panel:} The mass distribution of the subpopulation in the upper panel is represented by full bars, whereas empty bars illustrate the mass distribution of their progenitors.}
         \label{fig:numax2060}
   \end{figure}

\begin{figure}
   \centering
   \includegraphics[width=.91\hsize]{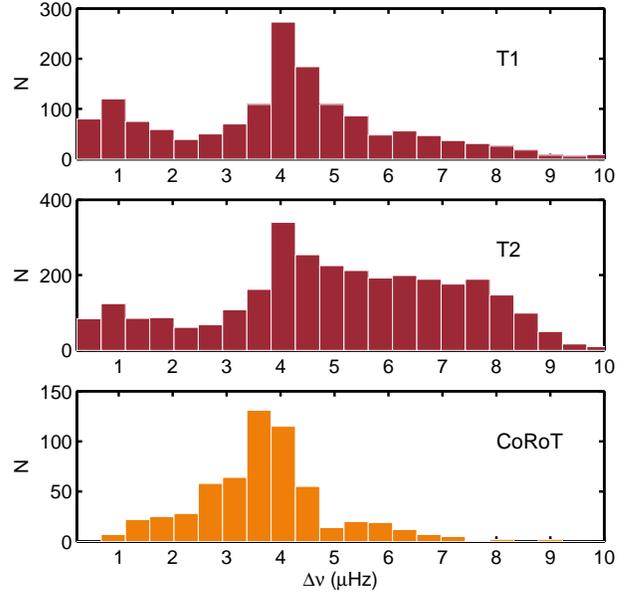}
   \caption{Histogram showing the comparison between the $\Delta\nu$ distribution of the observed (\emph{lower panel}) and simulated populations of red giants: T1 (\emph{upper panel}) and T2 (\emph{middle panel}).
              }
         \label{fig:dnu}
   \end{figure}

The comparison with the observed \numax\ distribution  suggests that the bulk of CoRoT LRc01 pulsating giants are red clump stars. In Fig. \ref{fig:numax2060}, upper panel, we show the location in the HR diagram of stars belonging to T1 and having 20 $\mu$Hz $< \nu_{\rm max} < 50$ $\mu$Hz. This lets us identify the dominant peak in the distribution as due to red-clump stars, i.e. low-mass stars with actual masses $0.8 \lesssim M/M_\odot \lesssim 1.8$ (see the lower panel of Fig. \ref{fig:numax2060}), which are in the core He-burning phase after having developed an electron-degenerate core during their ascent on the RGB and passed the He-flash.
This conclusion is also fully supported by the comparison between the observed and theoretical distribution of $\Delta\nu$, as shown in Fig. \ref{fig:dnu}.
   \begin{figure}
   \centering
   \includegraphics[width=.9\hsize]{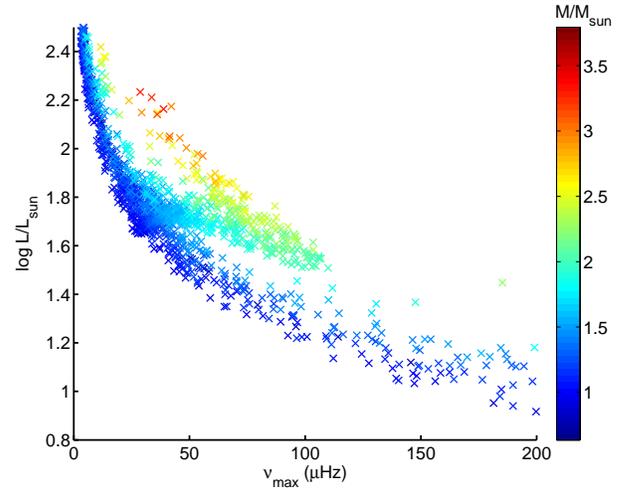}
   \caption{\numax\ vs. $\log{L/L_\odot}$ for each star in the T1 population. The mass of the star is colour-coded (see colour bar on the right).}
         \label{fig:numax_all}
   \end{figure}

A clearer relation between the \numax\ distribution and the properties of the population can be inferred from Fig. \ref{fig:numax_all}:\begin{itemize}
\item stars with \numax\ $\lesssim 20\, \mu$Hz are predominantly high-luminosity, low-mass stars ascending the giant branch;
\item in the \numax\ domain between 20 $\mu$Hz and 110 $\mu$Hz we can distinguish the dominant contribution of red-clump stars, burning helium at a nearly constant luminosity ($\log{L/L_\odot} \simeq 1.7$, and \numax\ in the 20-40 $\mu$Hz range), and that of stars with $1.8 \lesssim M/M_\odot \lesssim 2.5$. The latter populate the slightly less luminous ($1.5 \lesssim \log{L/L_\odot} \lesssim 1.8$) secondary clump \citep{Girardi99}, and have higher \numax\ ($40$ $\mu{\rm Hz}\lesssim$  \numax\ $\lesssim 110$ $\mu$Hz);
\item the very few stars with even higher masses ($M/M_\odot \gtrsim 2.5$) have larger radii and thus smaller \numax\ ($\lesssim 70$ $\mu$Hz);
\item only a negligible fraction of stars in the population have \numax\ $\gtrsim 110$ $\mu$Hz, a domain populated exclusively by low-mass, low-luminosity RGB stars.
\end{itemize}
This leads to the existence of a maximum value of \numax\ ($\simeq 110$ $\mu$Hz) for stars in the long-lived core-He-burning phase.

We note that the robustness of these results has been successfully checked by using different synthetic population models, such as {\sc basti} \citep{Pietrinferni04} and the Besan\c{c}on model of the Galaxy \citep{Robin03}.

\section{Discussion}
\label{sec:discu}
The observed dominant peaks in the \numax\ and $\Delta\nu$ distributions are qualitatively explained by the models. Nevertheless, a Kolmogorov-Smirnov test shows that the discrepancies between the simulated and observed distributions are, from a statistical point of view, highly significant because the samples contain a large number of stars.

As discussed in Sect. \ref{sec:pop}, the predicted \numax\ distribution is affected by several parameters in stellar populations models.
We note, however, that, since  most of the stars belonging to the RC cover a limited range in radius, \numax\ represents a valuable tracer of the mass distribution of the population (see Fig. \ref{fig:numax2060}).
The mass distribution of RC stars is, in turn, highly dependent on the age and, more generally, on the star formation rate of the composite population, as well as on the mass-loss rate adopted during the RGB.



As shown in Figs. \ref{fig:numax} and \ref{fig:dnu}, the position of the dominant peak in the observations is at a slightly lower frequency than in the theoretical \numax\ and $\Delta\nu$ distributions.
This discrepancy can be reduced by considering an older population or by increasing the star formation rate beyond 8 Gyr ago, which could also translate into a larger contribution of the 11-12 Gyr-old thick-disk population.
A further possibility for reconciling the observed and predicted \numax\ distribution is that mass loss during the RGB  is more efficient than accounted for in the models. In fact, while mass loss is needed to understand the observed properties of horizontal branch stars, there is no satisfactory physical description of this phenomenon or any general consensus on the empirical formula to describe it, in particular concerning its dependence on metallicity \citep[see for a recent review Sect. 5 of ][and references therein]{Catelan09}.

The behaviour of the observed \numax\ and $\Delta\nu$ distributions at values of \numax\ $\gtrsim 40$ $\mu$Hz and $\Delta\nu \gtrsim 5$ $\mu$Hz qualitatively agree with T1, where a constant star formation rate was adopted. However, various studies in the literature suggest that the local disk experienced a recent burst of star formation \citep[see e.g. ][]{Rocha00}. We thus consider a second synthetic population (T2) where we simulate a recent burst in {\sc trilegal} by adding to T1 a young population, obtained assuming a constant star formation rate between 0-1 Gyr. We find in T2 a significant increase in stars with relatively high \numax\ and $\Delta\nu$ (as shown in the middle panels of Figs. \ref{fig:numax} and \ref{fig:dnu}), caused by the increased population with $M \gtrsim 1.8$ $M_\odot$. This comparison suggests that a recent SF burst is not supported by the observed distributions, though a detailed analysis of observational biases at \numax$\gtrsim 40$ $\mu$Hz has to be carried out to make quantitative inferences on the recent star formation rate of the local disk \citep[see ][ for a discussion]{Hekker09}.

\section{Concluding remarks and future prospects}
We obtain a robust identification of the bulk of CoRoT LRc01 red giants as red clump stars. This conclusion is only based on seismic constraints, namely on the comparison between the \numax\ and $\Delta\nu$ distributions observed and those based on a synthetic stellar population. The latter was computed with the population synthesis code {\sc trilegal}, assuming standard parameters \citep[see ][]{Girardi05} to describe the morphology and the star formation history of the Galaxy components.

There is very good global agreement between the observed and simulated distributions of \numax\ and $\Delta\nu$. However, the dominant peak in the observations is located at a slightly lower frequency than our theoretical \numax\ and $\Delta\nu$ distributions, which suggests an underestimated fraction of low-mass stars. We find that possible solutions for a better agreement (i) increase the relative contribution of the old ($\gtrsim$ 8 Gyr) population or (ii) increase mass loss during RGB.
The behaviour at high frequencies can possibly rule out recent bursts of star formation (0-1 Gyr ago). To ascertain this point, a thorough discussion of possible observational biases at high frequencies is needed, however. We note that a test of the scalings of \numax\ and $\Delta\nu$  (Eqs. \ref{eq:numax} and \ref{eq:dnu}) in nearby well-constrained red-clump giants would also represent a major step in further validating the asteroseismic inferences. In this context, ground- and space-based observations of solar-like oscillations in bright RC stars play a complementary role in the study of red-giant populations.

As a further development of this study, we note that the asteroseismic identification of a large number of RC stars in the disk will also, when combined with additional photmetric/spectroscopic constraints, allow us to estimate the distances of the stars in the population, provided that metallicity/age effects on the absolute RC magnitude are taken into account \citep[see e.g.][ and references therein]{Salaris01} .

In this Letter we have shown that the detection of solar-like oscillations in populations of red giants with CoRoT allows us to access the properties of (so far) poorly-constrained populations.
Our view of the Galaxy and of its constituents promises to be enriched and sharpened by such constraints, and by the forthcoming observations of populations in different galactic fields with CoRoT and Kepler \citep{JCD08}.

\begin{acknowledgements}
The authors acknowledge L. Girardi, A. Pietrinferni, and S. Cassisi for their kind help with population synthesis codes.
\end{acknowledgements}

\bibliographystyle{aa}
\small
\bibliography{12822}
%
%

\end{document}